\documentclass[sigconf, nonacm]{acmart}

\newcommand\vldbdoi{10.14778/3554821.3554853}
\newcommand\vldbpages{3598 - 3601}
\newcommand\vldbvolume{15}
\newcommand\vldbissue{12}
\newcommand\vldbyear{2022}
\newcommand\vldbtitle{\shorttitle} 
\newcommand\vldbavailabilityurl{}
\newcommand\vldbpagestyle{empty}

\usepackage[noend]{algpseudocode}
\usepackage{algorithm}
\usepackage{bbm}
\usepackage{graphicx}
\usepackage{listings}
\usepackage[compress]{cleveref}
\usepackage{mathrsfs}
\usepackage{bm}
\usepackage{balance}
\usepackage{outlines}
\usepackage{svg}
\usepackage{grffile}
\usepackage[caption=false]{subfig}
\usepackage{xspace}
\usepackage{enumitem}
\usepackage{xcolor}
\usepackage{array}
\usepackage{nccmath}
\usepackage{multirow}
\usepackage{pifont}
\usepackage{soul}
\usepackage{caption}
\usepackage{upgreek}

\usepackage{xpatch}
\xpatchcmd{\NCC@ignorepar}{%
\abovedisplayskip\abovedisplayshortskip}
{%
\abovedisplayskip\abovedisplayshortskip%
\belowdisplayskip\belowdisplayshortskip}
{}{}

\graphicspath{./figures/}

\crefname{section}{Section}{Sections}
\Crefname{section}{Section}{Sections}
\crefname{figure}{Figure}{Figures}
\Crefname{figure}{Figure}{Figures}
\crefname{subfigure}{Figure}{Figures}
\Crefname{subfigure}{Figure}{Figures}
\crefname{algorithm}{Algorithm}{Algorithms}
\Crefname{algorithm}{Algorithm}{Algorithms}
\crefname{table}{Table}{Tables}
\Crefname{table}{Table}{Tables}
\crefrangelabelformat{subfigure}{#3#1#4--#5(\crefstripprefix{#1}{#2}#6}

\definecolor{emerald}{rgb}{0.31, 0.78, 0.47}

\newcommand{\eat}[1]{}

\newcommand{\stitle}[1]{\vspace{0.5ex}\noindent{\bf #1}}

\newcommand{\preeq}{\vspace{0mm}\begin{small}}
\newcommand{\posteq}{\vspace{0mm}\end{small}}

\newcommand{\system}{TQP\xspace}
\newcommand{\lsystem}{\textsc{Hummingbird}\xspace}

\usepackage{url}

\DeclareFixedFont{\ttb}{T1}{txtt}{bx}{n}{8}
\DeclareFixedFont{\ttm}{T1}{txtt}{m}{n}{8}

\usepackage{color}
\definecolor{deepblue}{rgb}{0,0,0.5}
\definecolor{deepred}{rgb}{0.6,0,0}
\definecolor{deepgreen}{rgb}{0,0.5,0}
\definecolor{purple}{rgb}{0.5,0,0.5}
\definecolor{gray}{rgb}{0.33,0.33,0.33}

\definecolor{dkgreen}{rgb}{0,0.6,0}
\definecolor{gray}{rgb}{0.5,0.5,0.5}
\definecolor{mauve}{rgb}{0.58,0,0.82}

\lstdefinelanguage{Python}{
	keywords={typeof, torch, nonzero, index_select, zeros_like, lt, masked_select, new, true, false, catch,def,val, function, return, null, catch, switch, var, shape,  while, do, else, case, break, override},
	keywordstyle=\color{blue}\bfseries,
	ndkeywords={class, export,extends, boolean, throw, implements, import, this, abstract, for, in, if},
	ndkeywordstyle=\color{dkgreen}\bfseries,
otherkeywords={+, =>,<=, ==, >,< , || , T},
	identifierstyle=\color{black},
	sensitive=false,
	comment=[l]{//},
	morecomment=[s]{/*}{*/},
	commentstyle=\color{purple}\ttfamily,
	stringstyle=\color{red}\ttfamily,
	morestring=[b]',
	morestring=[b]"
}
\author{Yuki Asada$^{*2}$, Victor Fu$^{*2}$, Apurva Gandhi$^{*1}$, Advitya Gemawat$^{*1}$, Lihao
Zhang$^{*2}$, Dong He$^{4}$, \\Vivek Gupta$^{3}$, Ehi Nosakhare$^{1}$, Dalitso Banda$^{1}$, Rathijit Sen$^{1}$, Matteo Interlandi$^{1}$}

\affiliation{
    \institution{$^{1,2,3}$Microsoft, $^4$University of Washington}
    \country{}
}
\affiliation{\{$^1$<firstname>.<lastname>,$^2$<firstname><lastname>,$^3$vivgupt\}@microsoft.com, $^4$donghe@cs.washington.edu \country{}}

\begin{document}

\title{Share the Tensor Tea: How Databases can Leverage the Machine Learning Ecosystem\vspace{-0.8ex}}
%\title{How Databases can Drink the Machine Learning Tea}
%\title{How Tensors Let Databases Drink the Machine Learning Tea}
% \title{TQP: How the Database World can Drink the Machine Learning Tea}
%\title{How to Catch Two Pigeons With One Fava Bean}
%\title{Demonstration of How to Save Two Birds with One Nest} 
%\title{Demonstration of How to Kill Two Birds with One Stone}

\begin{abstract}
\vspace{-0.5ex}
%\carlo{put some concrete number callouts}
%\jcr{first attempt}

\begin{sloppypar}

%In this demo paper %we will showcase how databases can take advantage of the innovations happening in the machine learning (ML) domain.
%Specifically, 
We demonstrate Tensor Query Processor (\system): a query processor that automatically compiles relational operators into tensor programs.
By leveraging tensor runtimes such as PyTorch, TQP is able to: (1) integrate with ML tools (e.g., Pandas for data ingestion, Tensorboard for visualization); (2) target different hardware (e.g., CPU, GPU) and software (e.g., browser) backends; and (3) end-to-end accelerate queries containing both relational and ML operators.
\system is generic enough to support the TPC-H benchmark, and it provides performance that is comparable to, and often better than, that of specialized CPU and GPU query processors.\vspace{-2ex}

% As a result, TQP is able to unlock the following existing machine learning features for databases
% \system is able to efficiently run the full TPC-H benchmark by implementing novel algorithms for executing relational operators on the specialized tensor routines provided by TCRs. 
% Meanwhile, \system can target various hardware while only requiring a fraction of the usual development effort. 
 
 \end{sloppypar}
\end{abstract}

\maketitle

\pagestyle{\vldbpagestyle}
\begingroup\small\noindent\raggedright\textbf{PVLDB Reference Format:}\\
Yuki Asada, Victor Fu, Apurva Gandhi, Advitya Gemawat, Lihao Zhang, Dong He, Vivek Gupta, Ehi Nosakhare, Dalitso Banda, Rathijit Sen, Matteo Interlandi. \vldbtitle. PVLDB, \vldbvolume(\vldbissue): \vldbpages, \vldbyear.\\
\href{https://doi.org/\vldbdoi}{doi:\vldbdoi}
\endgroup
\begingroup
\renewcommand\thefootnote{}\footnote{\noindent $^*$ Equal contribution.

\noindent This work is licensed under the Creative Commons BY-NC-ND 4.0 International License. Visit \url{https://creativecommons.org/licenses/by-nc-nd/4.0/} to view a copy of this license. 
For any use beyond those covered by this license, obtain permission by emailing \href{mailto:info@vldb.org}{info@vldb.org}. 
Copyright is held by the owner/author(s). Publication rights licensed to the VLDB Endowment. \\
\raggedright Proceedings of the VLDB Endowment, Vol. \vldbvolume, No. \vldbissue\ %
ISSN 2150-8097. \\
\href{https://doi.org/\vldbdoi}{doi:\vldbdoi} \\
}\addtocounter{footnote}{-1}\endgroup

\ifdefempty{\vldbavailabilityurl}{}{
\vspace{.3cm}
\begingroup\small\noindent\raggedright\textbf{PVLDB Artifact Availability:}\\
The source code, data, and/or other artifacts have been made available at \url{\vldbavailabilityurl}.
\endgroup
}

\begin{sloppypar}

\vspace{-1.5ex}
\section{Introduction}
\vspace{-0.5ex}
\label{sec:introduction}

The free lunch is over! While from 1985 to 2010 the CPU performance doubled approximately every 1.5 years,  
going forward the expectation is that CPU performance will only double every 20 years~\cite{dean2019deep}.
In this post-Moore's Law era, taking advantage of hardware acceleration is paramount. 
Both the database and the machine learning (ML) communities are well aware of this. 
In fact, in the last decade we have been witnessing a growing investment in techniques able to exploit  ``commodity'' accelerators such as SIMD instructions and GPGPUs. 
ML especially is becoming so predominant that many devices now come with ML-specific hardware accelerators (e.g., Apple Neural Engine, NVIDIA Tensor Cores), and the amount of money poured by venture capitalists on startups focusing on new hardware for ML is soaring~\cite{ai-hw-market}. 
This trend is mostly driven by the computation requirements of the state-of-the-art computer vision and NLP models. %which is increasing by approximately a factor of 15 every 2 years~\cite{ai-memory-wall}. 

 However,  programming hardware accelerators is notoriously hard.
 The ML community has a remarkable tool for
easing the authoring and the deployment of ML models: the \emph{tensor abstraction}~\cite{hummingbird-vision}.
Using this abstraction, ML practitioners can enjoy writing their models in high-level languages, while being free to run them on any hardware and backend of their choice: let it be CPUs, GPUs, custom ASICs, edge devices, or even browsers~\cite{ort-wasm}.
In the middle, ML frameworks such as PyTorch~\cite{pytorch} are able to map tensor programs into efficient executions over the target backends.
In this paper, we will refer to any ML framework supporting tensor programs as Tensor Computation Runtimes (TCRs).

Unfortunately, if we look at the analytical database space, we see a different picture. 
In the last several decades, the database community has been focusing on squeezing every last bit of performance from hardware devices (e.g., SIMD, GPUs), but, as a matter of fact, all these efforts are fragmented, specialized, and unrelated to each other, whereby users cannot enjoy the ease of deployment as well as the thriving ecosystem which makes Data Science (DS) and ML so popular.

In \cite{tqp-tr}, we showed how this gap can be filled thanks to Tensor Query Processor (TQP~\footnote{Pronounced `Teacup'.}): a SQL query processor built on top of TCRs. In this demonstration we will showcase \system features. 
Users of \system can: (1) take advantage of the flexibility and performance of TCRs, and execute their relational queries over any supported hardware (e.g., CPU, GPU, TPU) and software (e.g., browser) backends; (2) end-to-end accelerate queries containing mix of relational and ML operators such as prediction queries~\cite{predictsql}; and (3) seamlessly integrate with DS workflows and ML tools (e.g., TQP is \texttt{pip}-installable, and it leverages ML libraries such as Pandas and Tensorboard for data ingestion and visualization, respectively).
TQP is integrated with Apache Spark~\cite{spark} and PyTorch: it accepts input as a Spark SQL physical plan, and it uses a novel compilation stack based on~\cite{hummingbird} to lower relational operators into tensor programs implemented in PyTorch. %which can then be executed on different hardware.
%Alternatively, TQP allows exporting the tensor programs into different format and libraries (e.g., ONNX).
TQP is expressive enough to support all the 22 queries composing the TPC-H benchmark, and it often outperforms specialized CPU and GPU query processors. For example, on Q6 and Q14 at scale factor 1, TQP is more than 3$\times$ faster than Apache Spark on CPU, and more than 4$\times$ faster than BlazingSQL~\cite{blazing-sql} on GPU.
While TQP is still in the prototype phase, we are actively investigating how it can be leveraged by Microsoft products, e.g., through ONNX Runtime (ORT) integration~\cite{predictsql}. %We refer readers to our technical report for further details on TQP~\cite{tqp-tr}.

To showcase the versatility and performance of TQP, in this demo we will guide the audience through three different scenarios:

\begin{itemize}[style=multiline,wide,nosep]
\item \emph{In the first scenario, we will show how TQP can be integrated with DS tools and libraries} (Section~\ref{sec:scenario-tensorboard}). Specifically, we will showcase our integration with Tensorboard~\cite{tensorflow-white-paper}, and how this can be used for visualizing query plans (tensor programs), and profiling query characteristics as well as operator performance.
\item \emph{In the second scenario}, \emph{we will show how TQP can compile and execute TPC-H queries on different backends} (Section~\ref{sec:scenario-backend}). We will use an Azure VM equipped with an NVIDIA GPU device, and show the query performance over CPU, GPU, as well as browser execution on Web Assembly through ORT. 

\item \emph{In the third scenario, we will show how TQP can both express and end-to-end accelerate prediction queries} containing an ML model embedded into a SQL query  (Section~\ref{sec:scenario-ml}). Here we will show: (1) how we extended the Spark SQL syntax with a {\sc predict} keyword allowing us to execute, within SQL statements, both traditional ML (e.g., scikit-learn~\cite{scikit}) models and pre-trained neural networks; and (2) how relational operators and ML models are compiled into a unique tensor program that can be end-to-end executed on a GPU.
\end{itemize}

All the scenarios will be presented as a Python Notebook. Audience will be allowed to modify the queries and observe the performance tradeoffs; modify the ML models (Scenario 3), and interact with the UI (Scenario 1). 
\vspace{-1ex}
\section{System Overview}
In this section, we summarize TQP's data representation (Section~\ref{sec:data}), and query compilation and execution (Section~\ref{sec:compile}) previously introduced in~\cite{tqp-tr}.
%In this Section we briefly summarize how \system maps tabular data into tensors (Section~\ref{sec:data}), as well as how input SQL queries are compiled into tensor programs (Section~\ref{sec:compile}).
Finally, we report some experimental number on two TPC-H queries (Section~\ref{sec:experiments}). 
We refer readers to~\cite{tqp-tr} for more details on the system design, implementation, and algorithms mapping relational operators into tensor programs. 

\vspace{-1ex}
\subsection{Data Representation}
\label{sec:data}

%to the internal representation required by the PyTorch tensor programs it generates. 
\system internally represents tabular data in a columnar format. 
Each column of a table is represented as a $(n \times m)$ tensor, where $n$ is the input number of rows, and $m$ is the maximum length needed to store the column values. 
\textit{Numerical} columns are represented as $(n \times 1)$ tensors, i.e., a single vector column is enough to represent the numerical values. 
%Conversion of numerical columns into a tensor representation is often zero-copy. 
\textit{Date} columns are also represented as $(n \times 1)$ numeric tensors: i.e., the UNIX Epoch in nanoseconds. 
%However, in this case, serialization/deserialization may be required depending on the source/target date data representation. 
Finally, \textit{string} columns are harder to manipulate efficiently using tensors. 
Currently, \system represents string columns using a $(n \times m)$ numeric tensor, where $m$ is the maximum character length of any value for that column. 
To store a string value in the tensor, \system right-pads it with $0$s if its length is less than $m$. %then \system stores a character per tensor column. 
Currently, \system only supports UTF-8 encoded string values.
TQP automatically transform input data into the tensor format. Data transformation is in general zero-copy, except date and strings columns that require data conversion. %TQP supports as input tabular formats such as Pandas and Arrow. 

\vspace{-1ex}
\subsection{Query Compilation and Execution}
\label{sec:compile}

The TQP workflow for running a query is composed of two phases: in the \textit{compilation} phase, the input query is transformed into an executable program; in  the \textit{execution} phase, the program is run.
%Now  we briefly  summarize  each  phase. 
%Due to space limits here we only summarize each phase.

\stitle{Compilation.} TQP’s  compilation  phase is  composed  of  4  main  layers:  (1)  the \emph{parsing layer}  converts the  input  SQL  statements  into  an internal \textit{Intermediate  Representation (IR) plan} representing  the input query’s physical plan. The query physical plan is generated within TQP by an external frontend database system. (2) the \emph{optimization layer} implements a rule-based optimizer providing IR-to-IR transformations;  (3) the \emph{planning layer} translates  the  IR  plan  generated  in  the  previous  layer  into  an \textit{operator  plan} where  each  operator  in  the  IR  is  mapped  into  a tensor  program  implementation;  finally  (4)  the \emph{execution layer} generates an \emph{Executor} from  the  operator  plan.  The Executor  is  the  program  that  will  be  triggered at runtime on the selected target backend TCR and hardware.

In its current implementation, \system uses vanilla PyTorch as the implementation target of tensor programs in the planning layer. 
PyTorch programs are then lowered into different targets in the execution layer. Currently, PyTorch, TorchScript, ONNX, and TVM are the supported lower-level targets. On the frontend side, TQP currently support only Apache Spark, but we are planning to add support also for other databases. Since TQP translates physical  query  plans  into  an  IR,  the architecture decouples the physical plan specification from the other layers, therefore allowing to plug different frontends.
\system implements novel algorithms for expressing relational operators into tensor programs. %Our full paper~\cite{tqp-tr} contains details on  supported operators, and describes few important algorithms (e.g., join and aggregation).

\stitle{Execution.} TQP first converts tabular data into the tensor format (as described in Section~\ref{sec:data}), and then the converted data is fed  into the Executor program for generating the  query  result.

\vspace{-1ex}
\subsection{Performance Evaluation}
\label{sec:experiments}

We report a performance evaluation comparing TQP executing TPC-H queries 6 and 14 on CPU, GPU and web browser, versus Apache Spark (on CPU) at scale factor 1.   
These two queries will also be used through the demo. 
Note that the queries, albeit simple, are not trivial: query 6 contains four filters (over different columns data types), and aggregation. Query 14 contains a join, as well as aggregation over a {\sc case} expression containing a {\sc like} statement. For CPU and GPU, we instrument TQP to use the TorchScript backend. For the web backend, we internally convert the queries in ONNX and run them on ORT with Web Assembly support~\cite{ort-wasm} in a JavaScript environment.

\begin{figure}[h]
    \centering
    \vspace{-2ex}
    \includegraphics[trim={8ex 16ex 0 16ex},clip,width=0.45\textwidth]{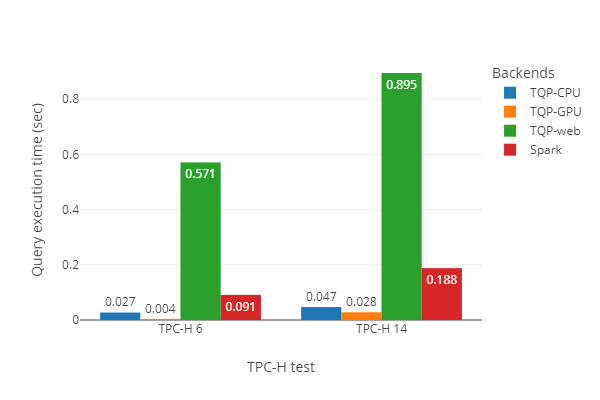}\vspace{-1.5ex}
    \caption{Query execution times for TPC-H 6  and 14 on Spark, and TQP on CPU, GPU, and web browser.}\vspace{-3ex}
    \label{fig:backends-bar}
\end{figure}

We use an Azure NC6 v2 machine equipped with 112 GB of RAM, an Intel Xeon CPU E5-2690 v4 @ 2.6GHz (6 virtual cores), and an NVIDIA P100 GPU (with 16 GB of memory). We run both TQP-CPU and Spark over all cores. We report the median of the execution time over 5 runs, after 5 warmup runs. The web backend runs on our personal laptop (a Surface Book 3) to mirror a common client-facing scenario. As shown in Figure \ref{fig:backends-bar}, TQP on CPU is around 3$\times$ faster than Apache Spark on both queries, while GPU execution is 20$\times$, and 6$\times$ faster, respectively. As expected, the web execution is quite slow. 
Besides the weaker computation power on the personal laptop, the performance of TQP on the web browser is depending on the efficiency of ORT, Web Assembly, and the JavaScript environment.
Nevertheless, this last result shows the flexibility of TQP, and its ability to leverage available open source ML tools.

%the performance of TQP on the web browser is also depending on the efficiency of ORT in the JavaScript environment. 
%\rs{Perhaps point out TQP makes browser execution possible? or else, is there a baseline to compare against?} \lz{added description of web browser execution in the first paragraph}

    \label{fig:backends}

\section{Demonstration}
In this Section we will introduce the three scenarios the TQP demo consists of.
Each scenario uses a Python Notebook run either locally on a laptop, or on an Azure VM with similar characteristics as the one described in Section~\ref{sec:experiments}. 

\begin{figure}[t!]
    \centering
    \includegraphics[trim={0 1ex 60ex 0},clip,width=0.43\textwidth]{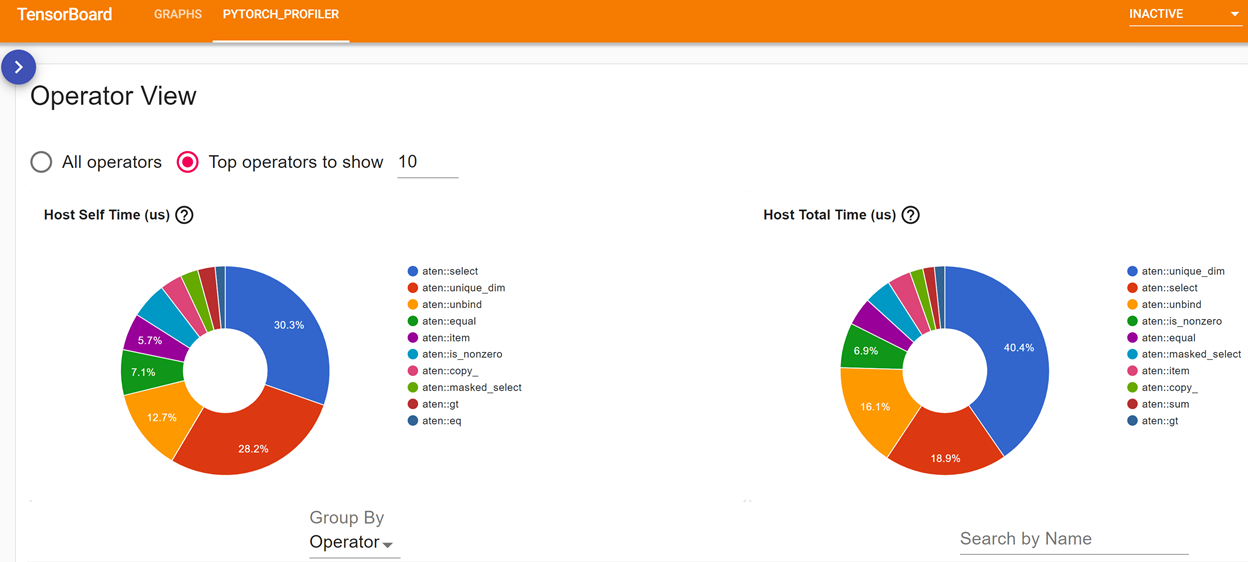}\vspace{-1ex}
    \caption{Runtime breakdown of the top operators for a selected query. The view is automatically generated within TQP  by the Pytorch Profiler, and visualized on TensorBoard.}\vspace{-4.5ex}
    \label{fig:operator_profile}
    \vspace{0.5ex}
\end{figure}

\vspace{-1ex}
\subsection{Scenario 1: Integration with DS Tools}
\label{sec:scenario-tensorboard}
% Image showing the pie charts with the operators performance.
TQP directly integrates with popular DS tools to offer a familiar developer experience. %and avoid reinventing the wheel.
For instance, TQP is \texttt{pip}-installable, it is fully implemented in Python, and it only depends on well-known, enterprise-grade libraries. TQP leverages Apache Spark (\texttt{pyspark}) to author their queries, and Pandas~\cite{pandas}, Numpy~\cite{numpy} and Apache Arrow~\cite{arrow} for data ingestion. 
%TQP integrates and expands \lsystem~\cite{hummingbird} to perform ML predictions over traditional ML models created by popular libraries such as scikit-learn \cite{scikit}, and it naturally support neural network models implemented in PyTorch.
Finally, generating tensor programs for query execution opens the door to utilizing ML tools for profiling relational operations.
In the first scenario, we will go over: (1) the integration with Apache Spark for plan generation; (2) integration with Pandas for data ingestion; and finally (3)
integration with Tensorboard~\cite{tensorflow-white-paper} for query profiling.

% In Scenario 3, we'll show how TQP can be complemented with Hummingbird \cite{hummingbird} to perform ML inference with a SQL query, which also allows users to plug-in traditional ML models created from popular libraries such as sklearn \cite{scikit}, LightGBM \cite{lgbm} and XGBoost \cite{xgboost}. % (which converts traditional ML models from popular libraries such as sklearn \cite{scikit} to a TCR model) to jointly perform data operations and ML inference with a single SQL query over TCRs.

%For queries executed on the pytorch \cite{pytorch} TCR, TQP integrates with TensorBoard \cite{tensorflow-white-paper} to visualize execution activity via the pytorch profiler. 
In this scenario, the audience will be guided through the following steps: (1) we will show how \system can be easily \texttt{pip}-installed and imported within a notebook; then
(2) we will import the {\sc lineitem} dataset scale factor 1 into the notebook
as a Pandas dataframe; (3) we will show how the selected queries can be compiled in TQP, and executed over the input dataframe; finally, (4) we will re-execute the query with the profiler activated and investigate runtime breakdowns of the query, execution trace, and memory utilization of the CPU/CUDA kernels in Tensorboard. 
Figure \ref{fig:operator_profile} highlights one of the runtime charts for TPC-H query 6. Moreover, leveraging TensorBoard's out-of-the-box capabilities also helps graphically visualize the generated tensor program of a query, as depicted in Figure~\ref{fig:execution_graph} for the query described in Scenario 3. 
%We will encourage audience to modify the query and interact with the visualizations. 

\begin{figure}[t!]
    \centering
    \includegraphics[trim={6ex 0 12ex 16ex},clip,width=0.48\textwidth]{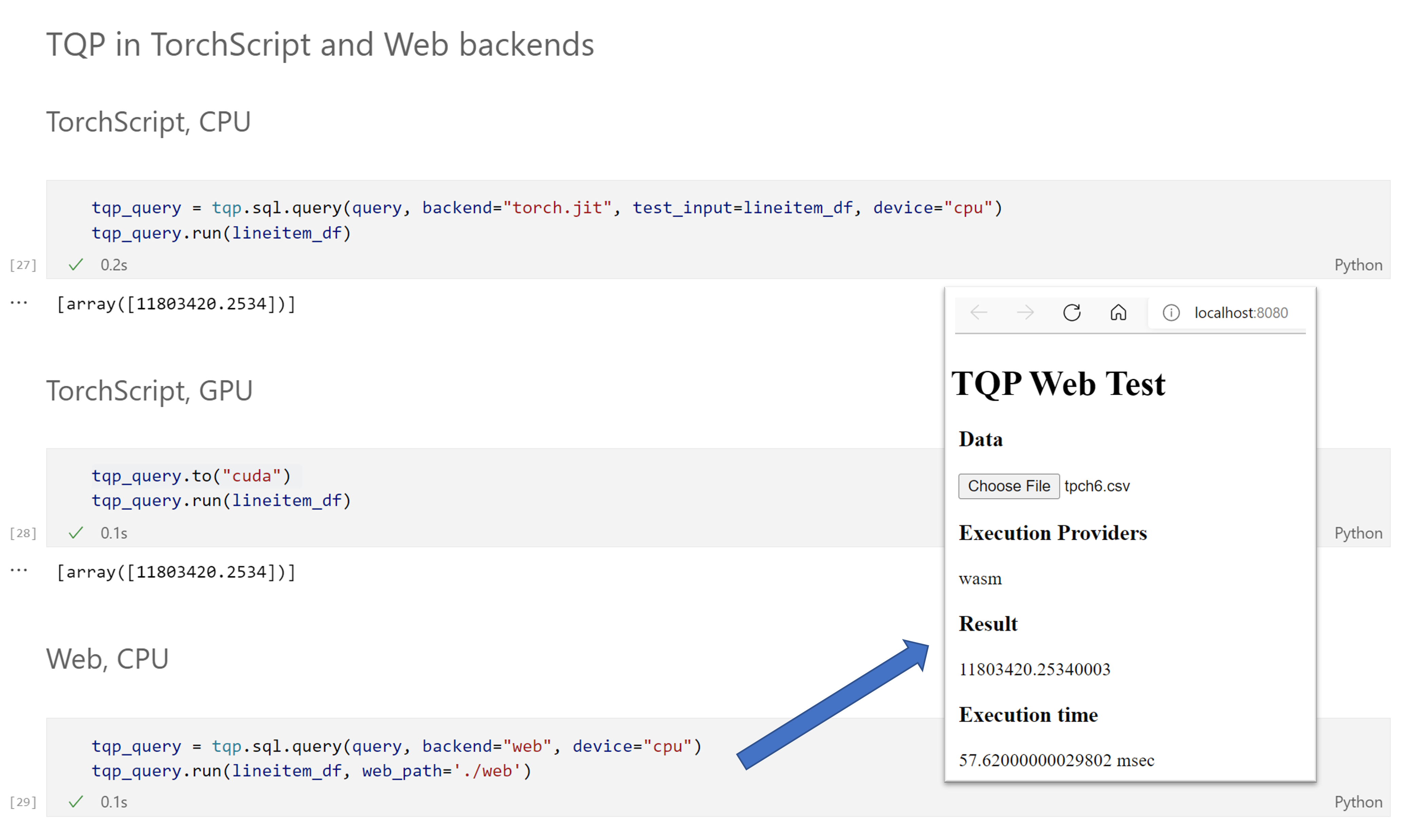}\vspace{-2ex}
    \caption{TPC-H query 6 executing in TQP on CPU and GPU devices using TorchScript (\texttt{torch.jit}) and over the \texttt{web} backend. Switching between different backends and hardware devices in TQP only needs one line of code change.}\vspace{-3.9ex}
        \label{fig:backends}
\end{figure}

\subsection{Scenario 2: Multiple Backend Support}
\label{sec:scenario-backend}

Extensive efforts have been made by the ML community to enable TCRs' compatibility with different software and hardware backends. For instance, PyTorch has full support for CPU and GPU. 
Furthermore, various ML formats like TorchScript and ONNX have been developed to improve the inference performance through optimizations, as well as enabling the deployment of models in non-Python environments such as edge applications or web browsers.
%For example, TorschScript is Pytorch intermediate representation allowing to serialize Pytorch models into C++ executable, while enabling optimizations such as operator fusion and constant folding.
%TQP design leverages PyTorch within its core compiler stack to implement tensor programs in the planning layer, naturally users can also utilize these frameworks to add extra optimization and compatibility in the later query execution. 

To showcase how relational queries can leverage the above features, in this scenario we will demo how TQP can compile and execute the two selected queries on: (1) the CPU using the TorchScript backend (\texttt{torch.jit}); (2) switching to a GPU device; and (3) on CPU on a web browser using the \texttt{web} backend. 
%(TQP in this case internally exports the Pytorch tensor program into ONNX, and execute it on a Javascript page using ONNX Runtime with Web Assembly support~\cite{ort-wasm}). 
~\footnote{Note that we could also run queries on the \texttt{web} backend targeting the GPU device through WebGL. However, the current implementation of ORT for WebGL does not cover all the ONNX operators, and hence execution currently fallback to CPU.} 
The workflow for this scenario is as follows: (1) initially we import all the required libraries; then (2) we import the {\sc lineitem} dataset scale factor 1 into the notebook as a Pandas dataframe; and (3) we show the TPC-H 6 and 14 queries. 
%using native PyTorch backend in a Python environment using both CPU and GPU. Also we will present the TQP can utilize ONNX runtime to execute the query in a demo web app.
Then, as presented in Figure \ref{fig:backends} for query 6, (4) we will show how TQP compiles the queries into tensor programs, and how
we can easily change the target device and the backend.
Finally, (5) we will run the input data through the compiled queries (or load the data into the web browser for the \texttt{web} backend), and show how all of them generate the same correct result. 
We also plan to compare the performance of TQP with Spark over the two TPC-H queries. Figure~\ref{fig:backends-bar} shows the expected results.

%\mi{Few comments: sk\_query -> tqp\_query. Can you please use the white background? Can you please add the parameter name for torch and web (I think it is backend?) Can you remove the SKIP\_NULLABLE thing from web (implementation detail). If it fits, can you please also add to the notebook a bar plot showing the different performance of the 3 backends?}

\vspace{-1.5ex}
\subsection{Scenario 3: Prediction Queries}
\label{sec:scenario-ml}
\vspace{-0.5ex}

\begin{figure*}[t!]
    \centering
    \includegraphics[width=0.9\textwidth]{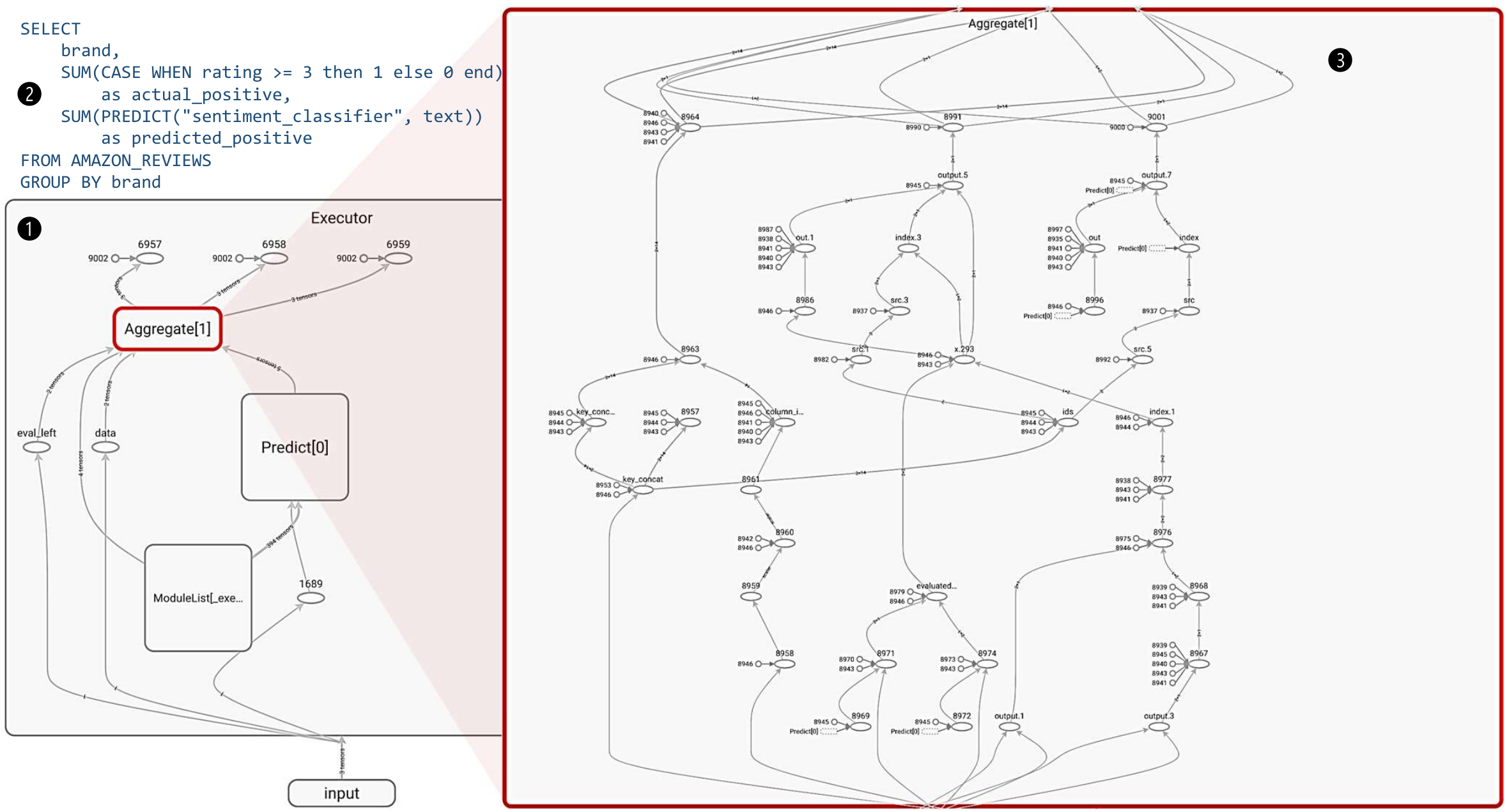}\vspace{-2ex}
    \caption{Executor graph for a prediction query (\ding{202}). The query (\ding{203}) combines an ML model, with a group-by-aggregate statement embedding a \textsc{case} expression. The query is used to predict the sentiment for each review, and compares the predictions with the user-provided ratings. The query-graph is interactive and the audience can further double-click on the various components to visualize how each operator is implemented as tensor operations (\ding{204} shows a zoom-in for the aggregation operator).}
    \vspace{-3ex}
    \label{fig:execution_graph}
\end{figure*}

Many past works have explored how to run ML workloads on relational engines (e.g., \cite{raven-sigmod, flock}). Doing so enables in-database ML scenarios with benefits such as improved performance by minimizing data movement, or reducing engineering effort by eliminating the need of a dedicated service for model predictions.
This latter feature is already available in commercial database systems. For example, since 2019, Microsoft SQL Server provides the ability to run in-database ML predictions through the \textsc{predict} keyword. 
This feature allows users to combine, in a single SQL query, model predictions and standard relational operations.
However, today such in-database ML integrations are typically done by spawning separate runtimes for relational and ML computations~\cite{raven-sigmod} (e.g., ORT in the SQL Server example). Given that TQP unifies relational
and ML runtimes, embedding ML into relational queries is even
more natural. %\apurva{Rephrase: Given that TQP unifies relational and ML runtimes, embedding ML into relational queries is even more natural.}
\footnotetext{The Executor graph used in Figure 4 can be accessed at
\url{https://tensorboard.dev/experiment/3mxfryX2QIaXWLVHlHkdIw/\#graphs}.}
To showcase this integration, in the third scenario we 
%TQP seamlessly combines machine learning model inference with relational operators. Particularly,
will describe our implementation of Microsoft SQL Server's \textsc{predict} keyword in TQP. 
% TQP integrates and expands \lsystem~\cite{hummingbird} to perform ML predictions over traditional ML models created by popular libraries such as scikit-learn \cite{scikit}, and it naturally support neural network models implemented in PyTorch.
Since TQP uses PyTorch as its default TCR backend, TQP natively supports predictions over any models built using PyTorch. Furthermore, since TQP integrates and expands \lsystem~\cite{hummingbird}, predictions over traditional ML models (e.g., created by libraries such as scikit-learn \cite{scikit}) are supported as well. 
Finally, hybrid ML-SQL queries too profit from all the features described in the previous scenarios, e.g., end-to-end acceleration on GPU and integration with DS tools such as TensorBoard.

In this demo scenario, we will guide the audience through the creation and execution of predictive queries for two tasks: (1) Sentiment Classification on the Amazon Product Reviews Dataset \cite{amazon_consumer_reviews}; and (2) Regression on the Iris Dataset \cite{iris}. The audience will be able to try a variety of models ranging from state-of-the-art, pre-trained transformers models from the HuggingFace Transformers library~\cite{transformers}, to traditional ML models such as those available in the scikit-learn library. We will also show how to combine these models with relational operations such as filters or aggregates in a single SQL query.
Figure~\ref{fig:execution_graph} shows an example of such a query, and its execution graph visualized in the TensorBoard tool.\footnotemark 

\vspace{-2ex}
\begin{acks}
\vspace{-0.5ex}
We would like to thank Magdalena Balazinska, Soundar Srinivasan, Konstantinos Karanasos, Jes\'us Camacho-Rodrigues, Carlo Curino, and Raghu Ramakrishnan for their insightful feedback and support.
\end{acks}

%The Amazon Product Reviews is a dataset consisting of consumer review text and ratings for various Amazon products. We will demonstrate how we can use state-of-the-art, pre-trained transformer models such as BERT model variants implemented in pure PyTorch by HuggingFace's transformers library for model inference such as sentiment classification over the product review text using SQL query syntax. The Iris dataset consists of measurements and types of numerous iris flowers. We will demonstrate 
%Image showing the graph with ml operators and relational one.

\end{sloppypar}

%\pagebreak
%\balance
\vspace{-2ex}
\bibliographystyle{ACM-Reference-Format}
\bibliography{main}

% \appendix
% \input{appendix}

\end{document}